\documentclass[pre,twocolumn]{revtex4}

	\usepackage{amsmath}
	\usepackage{makeidx}
	\usepackage{amsfonts}
	\usepackage{color}
	\usepackage[ansinew]{inputenc}
	\usepackage[usenames,dvipsnames]{pstricks}
	\usepackage{subfigure}
	\usepackage{epsfig}
	\usepackage{pst-grad} 
	\usepackage{pst-plot} 
	\usepackage[colorlinks,hyperindex]{hyperref}
	\usepackage{tikz,siunitx,mwe}%
	\hypersetup
	{
		colorlinks,%
		citecolor=black,%
		linkcolor=black,%
		urlcolor=black,%
	}

\begin{document}
\title{Electrical Autonomous Brownian Gyrator}

\author{K.-H. Chiang,
C.-L. Lee\footnote{E-mail: chilun@cc.ncu.edu.tw},
P.-Y. Lai,
Y.-F. Chen\footnote{E-mail: yfuchen@ncu.edu.tw}
}

\affiliation{Department of Physics, National Central University, Zhongli 32001, Taiwan}

\date{\today}

\begin{abstract}
We study experimentally and theoretically the steady-state dynamics of a simple stochastic electronic system featuring two resistor-capacitor circuits coupled by a third capacitor.  The resistors are subject to thermal noises at real temperatures.  The voltage fluctuation across each resistor can be compared to a one-dimensional Brownian motion.  However, the collective dynamical behavior, when the resistors are subject to distinct thermal baths, is identical to that of a Brownian gyrator, as first proposed by R. Filliger and P. Reimann in Physical Review Letters 99, 230602 (2007).  The average gyrating dynamics is originated from the absence of detailed balance due to unequal thermal baths.  We look into the details of this stochastic gyrating dynamics, its dependences on the temperature difference and coupling strength, and the mechanism of heat transfer through this simple electronic circuit.  Our work affirms the general principle and the possibility of a Brownian ratchet working near room temperature scale.
\end{abstract}


\maketitle

\section{\label{sec1}Introduction}
Richard Feynman explored whether one can extract work simply from the stochastic motions agitated by a surrounding heat reservoir in his famous discussion of the Brownian ratchet \cite{Feynman63}.  He pointed out that in order to extract work autonomously, it is necessary to have the system in contact with an additional cooling reservoir.  Therefore the second law of thermodynamics is demonstrated even under the consideration of microscopic stochastic dynamics.

Inspired by Feynman's discussion, and thanks to the recent advances in the manipulations on small-scale systems \cite{Martinez17}, a huge interest is emerging in the development of miniature thermal engines
 \cite{Reimann02,Hanggi09,VanDenBroeck04,Steeneken11,Blickle12,Martines16,Koski16,Davis01}.
In these systems with few degrees of freedom, one expects to extract work from the microscopic Brownian movements, and in contrast to bulk systems, their dynamics exhibits a prominent stochastic nature.  Among these studies, autonomous gyrators are more reminiscent of Feynman's original work.  For example, R. Filliger and P. Reimann \cite{Filliger07} introduced a Brownian gyrator, in which a structureless particle is simultaneously exposed to two heat reservoirs, each imposing on one of its motional degrees of freedom.  An average gyrating motion can be observed in the nonequilibrium steady state (NESS) \cite{Kwon05,Ghanta17} if the two reservoirs are of distinct temperatures, and hence this two-dimensional Brownian gyrator can serve as a ``minimal'' version of autonomous heat engines.  In Ref.~\cite{Filliger07}, the authors listed two more criterions for the generation of a Brownian gyrator: (1) the landscape for the confining potential is not rotationally symmetric, and (2) the directions along which the two random forces are imposed do not coincide with the principal axes of the potential landscape.

Despite the simple requirements in the Brownian gyrator or similar autonomous heat engines, it is technically challenging to expose a minuscule to two heat baths simultaneously, each upon an independent direction \cite{Seifert12}.  Experimentally, mechanical and electrical realizations of these autonomous engines are crafted with artificial thermal noises and are often featured by nonlinear interactions \cite{Eshuis10,Peng16,Serra-Garcia16,Hartmann15}.  Meanwhile, mesoscopic conductors with asymmetric and nonlinear couplings to multiple charge heat baths were proposed to generate unidirectional charge current \cite{Sanchez11,Sothmann12,Whitney16}.  Net electrical currents have been observed experimentally in those mesoscopic ratchets \cite{Roche15, Thierschmann15}, demonstrating their capability of rectifying heat to work.

In the current work, we report our studies on the stochastic dynamics of a capacitively-coupled resistor-capacitor ({\it RC}) circuit \cite{Ciliberto13, Ghanta17} as illustrated in Fig.~\ref{fig:figure1}(a)  , where the resistors are agitated by {\it real} thermal noises of two real heat baths.  We demonstrate that this simple linear system can be compared exactly to the Brownian gyrator as depicted in Ref.~\cite{Filliger07}.  In contrast to the existing studies on this circuit system concerning its entropy fluctuation and the applicability of fluctuation theorems \cite{Ciliberto13, Chiang17}, we turn our attention to its analogous gyrating behavior, which is concealed in its fluctuating dynamics over the configuration space.  While the voltage for each of the electrical element fluctuates due to thermal noises, an average heat is conducted from the hot to the cold reservoir via the circuit.  Along with the average unidirectional gyrating motion, they both are representations of the second law of thermodynamics.

The rest of the paper is organized as follows.  The experimental setup and its corresponding stochastic dynamical equation is given in Sec.~\ref{sec:setup}.  The main results concerning the gyrating dynamics over the configuration space, featuring its steady-state probability and flux distributions, are reported in Sec.~\ref{sssec:exp}.  In Sec.~\ref{sssec:theory}, we provide more theoretical details for gyrating dynamics of the NESS, and moreover, we also demonstrate the dependence between the gyrating direction and the temperature gradient.  In Sec.~\ref{sec:cycle} the dependence of energy flow and closed-cycle gyration is discussed with the aid of semi-adiabatic processes.  The dependencies of the rotating speed on the coupling strength and the temperature gradient are studied in Sec.~\ref{sec:rotation_speed}.

\section{\label{sec:setup}Experimental system}

\begin{figure}
\includegraphics[width=85 mm]{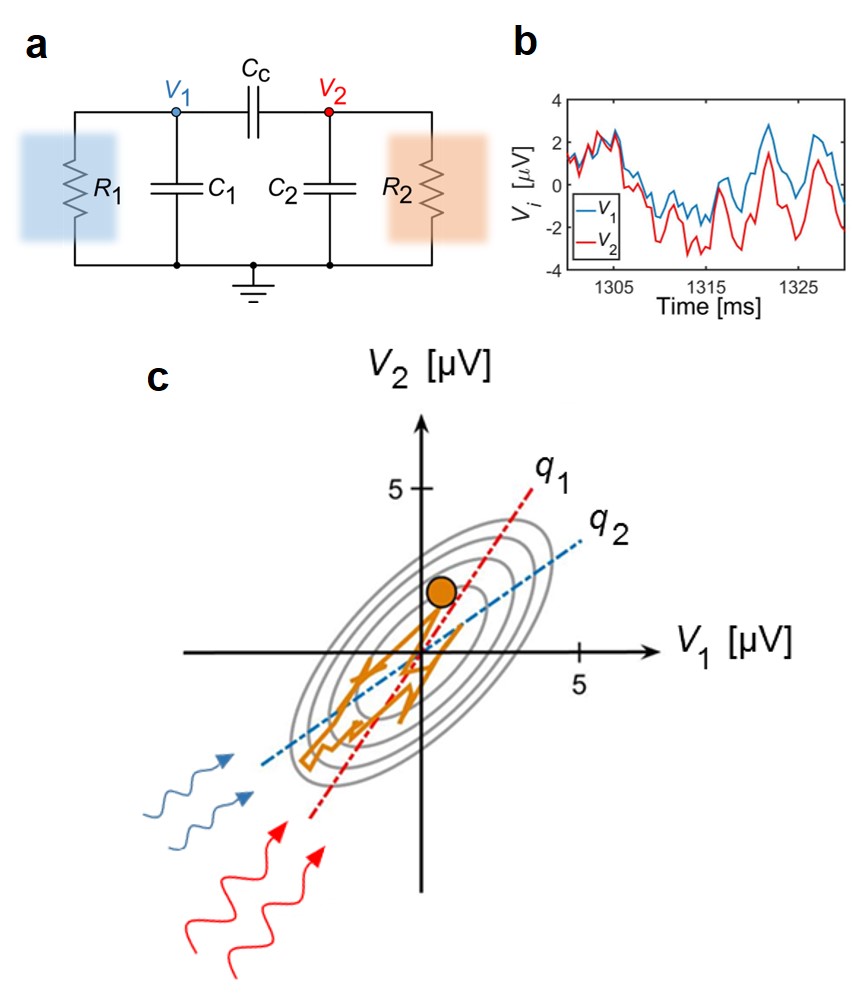}
\caption{Electrical autonomous Brownian gyrator.  (a) Schematics of the experimental system, featuring a capacitively-coupled {\it RC} circuit agitated by two heat baths.  (b) A snapshot of concurrent $V_1(t)$ and $V_2(t)$ over 30 ms with $C_c$ = 1.0 nF and $T_1$ = 120 K.  (c) A virtual particle evolving in the 2D phase space formed by $V_1$ and $V_2$.  A small segment of its trajectory (corresponding to the data in (b)) is shown by the orange line. The dashed lines indicate the $q_1$ and $q_2$ axes; see text.  The virtual particle is influenced by two heat baths and experiences two random noises $\xi_1$ and $\xi_2$ from directions parallel to $q_1$ and $q_2$ axes, respectively, as noises are depicted by two sets of wavy arrows.  The tilt ellipses designate potential contours with a minimum in the origin.}
\endcenter
\label{fig:figure1}
\end{figure}

Figure~\ref{fig:figure1}(a) shows the schematic of our study system.  Two {\it RC} circuits $(R_1,C_1)$ and $(R_2,C_2)$ are connected through a coupling capacitor $C_c$ \cite{Ciliberto13}.  The two resistors $R_1$ and $R_2$ are individually thermalized by the heat baths of temperature $T_1$ and $T_2$, respectively.  In our system of interest, the effects of electromagnetic induction are negligible, and the dynamics of the voltages across the resistors, $V_1(t)$ and $V_2(t)$, is governed by the coupled Langevin equation \cite{Ciliberto13}
\begin{equation}
  \hat{\bf R} \hat{\bf C} \dot{\vec{V}} = - \vec{V} + \vec{\xi}\, ,
\label{eq:equation1}
\end{equation}
where $\displaystyle \vec{V} \equiv \left( \begin{array}{c} V_1 \\ V_2 \end{array} \right)$,
$\vec{\xi} \equiv \left( \begin{array}{c} \xi_1 \\ \xi_2 \end{array} \right)$,
$\displaystyle \hat{\bf R} \equiv \left( \begin{array}{cc} R_1 & 0 \\ 0 & R_2 \end{array} \right) $, and
$\displaystyle \hat{\bf C} \equiv \left( \begin{array}{cc} C_1 + C_c & -C_c \\ -C_c & C_2 + C_c \end{array} \right) $.
The thermal (Johnson-Nyquist) noises $\xi_1$ and $\xi_2$ are Gaussian white and uncorrelated, namely $\langle \xi_i(t)\xi_j(t')\rangle=2k_BT_iR_i\delta_{ij}\delta(t-t')$, and $k_B$ is the Boltzmann constant.  Owing to a nonzero $C_c$, the dynamics of each voltage signal $V_i$ is influenced explicitly by both thermal noises.

The measured {\it RC} circuits in metal shielding boxes are placed in a
Faraday cage on an optical table.  The resistor $R_1$ in a metal shielding box is
cooled in a semiclosed liquid nitrogen dewar by liquid nitrogen vapor to create the
NESS.  We use voltage amplifiers with gain of $10^4$ to magnify the thermal voltages of
$V_1 and V_2$ before sampling.  The amplified signals are filtered by a 160-kHz
antialiasing filter, digitized at 262.1 kHz, and averaged over 128 digitized points for
a sample to achieve sampling rate of 2048 Hz.  Typically $10^6$ pairs of $(V_1, V_2)$
are recorded during each run.  The circuit parameters $C_1 = 488$ pF, $R_1 = 9.01$
M$\Omega$, $C_2=420$ pF, and $R_2=9.51$ M$\Omega$ are determined from the measured
noise power spectrums of $V_1$ and $V_2$ at $C_c = 0$ when both circuits are at room
temperature  (please refer to Ref.~\cite{Chiang17} for more experimental details and
characterization).  The coupling capacitance $C_c$ varies from 100 pF to 10 nF.  Its
value is independently obtained by a LCR meter.  The second reservoir is kept at room
temperature ($T_2=$ 296 K), and $T_1$ varies from 120 K to 296 K.  The value of $T_1$
below room temperature is measured by a K-type thermocouple, and can be re-affirmed by
the variance of statistics in $V_1$ with the knowledge of other circuit parameters.
Fig.~\ref{fig:figure1}(b) shows a snapshot of the concurrent voltage time traces
$V_1(t)$ and $V_2(t)$ with $C_c$ = 1.0 nF and $T_1$ = 120 K.  $V_1(t)$ and $V_2(t)$
resemble each other owing to the large $C_c$.

One can compare the electric circuit system to a Brownian particle in two dimensions, as depicted in Fig.~\ref{fig:figure1}(c).  The vector $\vec{V}(t)$ can serve as the position of this virtual Brownian particle at time $t$.  A small segment of its trajectory corresponding to the voltage time traces in Fig.~\ref{fig:figure1}(b) is shown by the orange line.  In the thermally equilibriated case, $T_1 = T_2$, the virtual particle does not exhibit any net movement besides thermal fluctuations; when $T_1 \neq T_2$, the particle is unequally agitated by the two heat baths, causing a persistent, unidirectional movement on average.  In the latter case, we set $T_1$ to be the colder heat bath throughout our study.

 \begin{figure*}
\includegraphics[width=170mm]{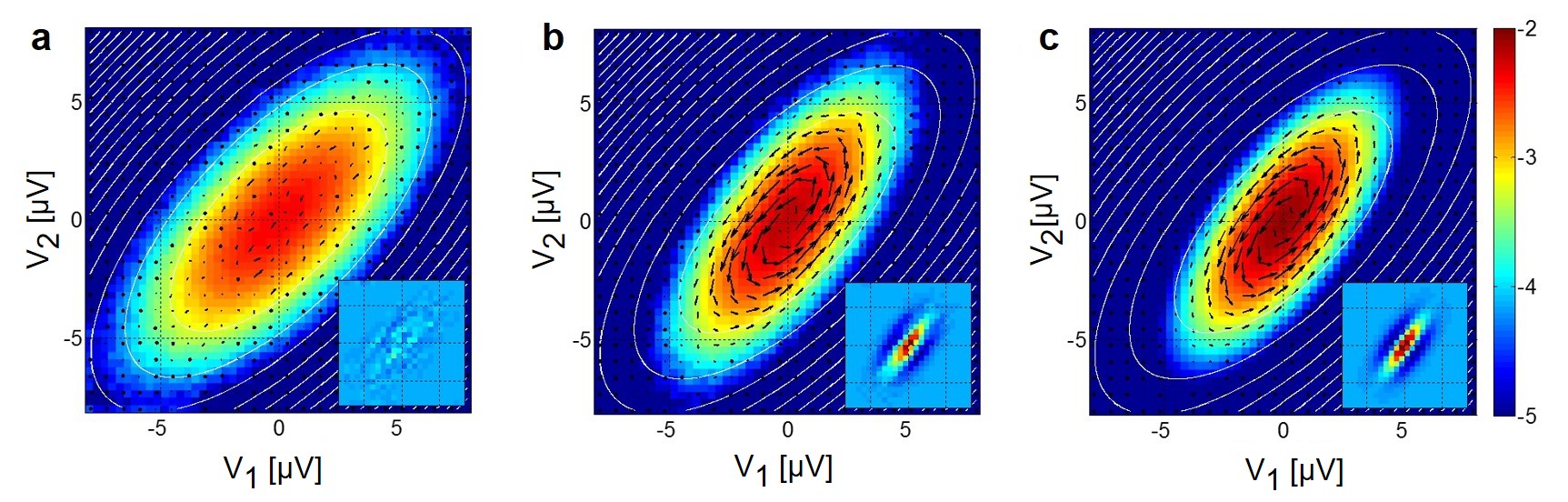}
\caption{Behavior of Brownian gyrator.  The figures present the main results of this work.  The value of $C_c=$ 1.0 nF is used here.  {\bf White contour -} the equipotential contours of the coupled {\it RC} circuit.  {\bf Colormap -} the steady state distribution ${P_{\rm ss}(\vec{V})}$.  {\bf Vector field -} the probability flux, $\vec{J}_{\rm ss}(\vec{V})$.  The inset colormap shows the curl of the probability flux, $\nabla \times \vec{J}_{\rm ss}$.  The experimental results are listed in (a) and (b).  (a) Equilibrium case ($T_1=T_2=296$ K).  The contour lines in $P_{\rm ss}$ and $U$ mutually agree, and $\vec{J}_{\rm ss}$ hardly exhibits any flowing trend as the detailed balance is valid.  (b) NESS case ($T_1=120$ K).  Contour lines of $P_{\rm ss}$ are tilted with respect to those in $U$, and $\vec{J}_{\rm ss}$ reveals a circulating trend about the origin.  (c) Theoretical counterparts of (b).}
\label{fig:figure2}
\end{figure*}

To compare our system to the Brownian gyrator described in Ref.~\cite{Filliger07}, we introduce the linear transformation
\begin{equation}
  \vec{q} \equiv \left( \begin{array}{c} q_1 \\ q_2 \end{array} \right) = \hat{\bf C} \vec{V}\, ,
\label{eq:equation2}
\end{equation}
where $q_1$ and $q_2$ represent the total capacitor charges in the neighbors of nodes 1 and 2, respectively.  The potential energy of the system (depicted by equipotential elliptical contours in Fig.~\ref{fig:figure1}c), as stored in the capacitors, is $U = \frac{1}{2}C_1V_1^2 + \frac{1}{2}C_2V_2^2 + \frac{1}{2}C_c(V_1-V_2)^2 = \frac12 \vec{V}^T \hat{\bf C} \vec{V} = \frac12 \vec{q}^{\,T} \hat{\bf C}^{-1} \vec{q}$.  With this transformation of variables, the coupled Langevin equation (Eq. (\ref{eq:equation1})) now reads
\begin{equation}
  \hat{\bf R} \dot{\vec{q}} = - \hat{\bf C}^{-1} \vec{q} + \vec{\xi} = - {\nabla}_q U + \vec{\xi}\, .
\label{eq:equation3}
\end{equation}
which is identical to the overdamped Langevin equation (Eq.(2)) in Ref.~\cite{Filliger07}.  Therefore, one can apply the framework of a Brownian particle confined in a 2D potential in Ref.~\cite{Filliger07} to the coupled {\it RC} circuit discussed in this work.  Note that the virtual particle simultaneously experiences two thermal noises $\xi_1$ and $\xi_2$ from directions parallel to the $q_1$ and $q_2$ axes, respectively, as noises are depicted by the wavy arrows in Fig.~\ref{fig:figure1}(c).  In this work, however, most results are presented on the $V_1-V_2$ plane (the physically observed variables), and the dynamics of each $V_i$ is influenced explicitly by both thermal noises.

\section{Behavior in nonequilibrium steady state: probability flux circulation}
\label{sec:flux_gyration}

\subsection{Experimental results}
\label{sssec:exp}

The main results of this work are visualized in Fig.~\ref{fig:figure2}. First, we present the equipotential lines of $U$ as white concentric elliptical contours in Fig.~\ref{fig:figure2}.  Due to the presence of a nonzero $C_c$, the potential possesses no rotational symmetry, and the principal axes of the contours are tilted in the $V_1-V_2$ frame and do not coincide with $q_1-q_2$ axes, where $\xi_1$ and $\xi_2$ act on.  Therefore, this setup meets the two aforementioned requirements and can serve as an electrical version of Brownian gyrator \cite{Filliger07}.

Our measured steady state distribution $P_{\rm ss}(\vec{V})$ is presented as a colormap plot in Figs.~\ref{fig:figure2}(a) and ~\ref{fig:figure2}(b).  At thermal equilibrium ($T_1=T_2 \equiv T$), $P_{\rm ss}$ follows the Boltzmann distribution.  Thus $P_{\rm ss}$ is constant on equipotential contours, as shown in Fig.~\ref{fig:figure2}(a).  The system falls into a NESS when $T_1 < T_2$ with a nonzero heat flow on average going from the $T_2$ heat bath to the $T_1$ heat bath through the circuit \cite{Ciliberto13}.  A NESS case of $T_1=$ 120 K is demonstrated in Fig.~\ref{fig:figure2}(b).  While $P_{\rm ss}$ in Fig.~\ref{fig:figure2}(b) still has an elliptic shape, it does not stay in accordance with the potential landscape.  Its principal axes rotate counterclockwise slightly when compared with those in Fig.~\ref{fig:figure2}(a).  This behavior is attributed to the narrower distribution in $V_1$ in the NESS case due to the lower $T_1$.

A major difference between thermal equilibrium and NESS lies in time reversibility.  Theoretically, the former is achieved through the detailed balance condition, which is itself a signature of time reversibility.  On the other hand, the detailed balance condition can fail in a nonequilibrium process, leading to persistent probability flows even in its steady state.  Here we evaluate the probability flux $\vec{J}_{\rm ss}(\vec{V}) \equiv P_{\rm ss}(\vec{V}) \vec{v}_{\rm flow}(\vec{V})$ from the experimental trajectory of the virtual particle, where $\vec{v}_{\rm flow}(\vec{V})$ represents the steady-state flow velocity at $\vec{V}$ \cite{Seifert12}.  We use the operational definition $\vec{v}_{\rm flow}(\vec{V}) \equiv (\langle\vec{V}(t+\Delta t)-\vec{V}(t)|\vec{V}(t)=\vec{V}\rangle - \langle\vec{V}(t)-\vec{V}(t-\Delta t)|\vec{V}(t)=\vec{V}\rangle)/2\Delta t$, where $\Delta t$ = 0.488 ms is the sampling interval (corresponding to the sampling rate of 2048 Hz), and the phase space is divided into grids with resolution $\Delta V = 0.67 \rm {\mu V}$ in order to accumulate decent statistics. The experimental results of $\vec{J}_{\rm ss}$ are presented as vector fields in Figs.~\ref{fig:figure2}(a) and~\ref{fig:figure2}(b).  There is clearly a circulating probability flux field in the NESS case (Fig.~\ref{fig:figure2}(b)), while no significant flow occurs in the thermal equilibrium case (Fig.~\ref{fig:figure2}(a)). Therefore, in a NESS, the motion of the virtual particle can be depicted by Brownian dynamics with an average counterclockwise circulation, i.e., it manifests as a Brownian gyrator in an electrical system. 

The circulation of probability flux in the NESS results from the unbalanced competition between the conservative and diffusive driving forces.  Na\"ively speaking, on the $V_1-V_2$ plane, the conservative force pulls the virtual particle inward.  Meanwhile, the diffusive force resulted from gradient changes of $P_{\rm ss}$ tends to push the virtual particle outward.  At thermal equilibrium (see Fig.~\ref{fig:figure2}(a)), the two sets of contour lines have identical shape, and their representative drives cancel out exactly.  Thus the net flux is zero everywhere, a signature of the detailed balance.  In the NESS case, however, due to the temperature difference, the principal axes for the contours of $P_{\rm ss}$ and $U$ are different, and the two driving forces are mostly not balanced.  As a result, the net force contributes to a non-vanishing flux. Circulating motion therefore emerges naturally since the flux at the steady state must be divergence-free (the curl of a nonzero field must exist somewhere for the divergence-free case).

Moreover, owing to the conservation of probability,
\begin{equation}
  \frac{dP_{\rm ss}}{dt} = \nabla P_{\rm ss} \cdot \vec{v}_{\rm flow} +\frac{\partial P_{\rm ss}}{\partial t} = 0
\end{equation}
holds along the steady-state circulation trajectories.  Since $\displaystyle \frac{\partial P_{\rm ss}}{\partial t} = 0$, $\vec{v}_{\rm flow}$ and thus $\vec{J}_{\rm ss}$ must be perpendicular to $\nabla P_{\rm ss}$~\cite{JpGP}.

The results for the curl of the steady-state flux, $\nabla \times \vec{J}_{\rm ss}$, are shown in the insets of Fig.~\ref{fig:figure2},  where $\nabla \times \vec{J}_{\rm ss}$ points out of the $V_1-V_2$ plane.  In the NESS case, the large positive curling trend near the origin causes the virtual particle to gyrate counterclockwise on average.  Note that away from the origin, small regions with a negative curling direction (shown as dark blue) can be observed from both our experimental and theoretical analysis.  Negative curl exists in regions of approximately parallel field lines whose magnitude decreases as the virtual particle marches outward.  Note that even in the negative curl regions, the flux field lines still follow the counterclockwise gyrating trend with respect to the origin.


\subsection{Theoretical analysis}
\label{sssec:theory}
We first consider
the corresponding Fokker-Planck equation of this stochastic system:
\begin {equation}
\begin {split}
  \frac{\partial P(\vec{V}, t)}{\partial t} =
  & \nabla \cdot [\hat{\bf{M}}^{-1} \vec{V} P(\vec{V},t)]\\
  &+\frac12 \nabla \cdot \hat{\bf{M}}^{-1} \hat{\bf{\Gamma}} (\hat{\bf{M}}^{-1})^{T} \nabla P(\vec{V},t)\, ,
\label{eq:FP}
\end {split}
\end {equation}

  where $\hat{\bf{M}} \equiv \hat{\bf R} \hat{\bf C} $, $\hat{\bf{\Gamma}}\equiv
  \left(\begin{array}{cc} \Gamma_1 & 0 \\ 0 & \Gamma_2 \end{array} \right) = \hat{\bf R} \hat{\bf T}$ and $\hat{\bf T} \equiv \left(\begin{array}{cc} 2k_B T_1 & 0 \\ 0 & 2k_B T_2 \end{array} \right)$ . 
It has a Gaussian steady-state distribution in
\begin{equation}
  P_{\rm{ss}}(\vec{V}) = \sqrt{\frac{\det (\hat{\bf{M}}^T \hat{\bf X} \hat{\bf{M}})}{\pi^2}} \exp(- \vec{V}^T \hat{\bf{M}}^T \hat{\bf X} \hat{\bf{M}} \vec{V})\, ,
\label{eq:ss_allV}
\end{equation}
where
\begin{equation}
  \hat{\bf X} \equiv \frac{ \displaystyle  (\hat{\bf M}^{-1})^T \hat{\bf \Gamma}^{-1} \hat{\bf M}^{-1} + \frac{\hat{\bf \Gamma}^{-1}}{\det(\hat{\bf M})} }
  { \displaystyle  \frac{ {\rm Tr}(\hat{\bf M}) }{\det(\hat{\bf M})} \left[ 1 + \frac{1}{\det(\hat{\bf \Gamma})}
  \left( \frac{ \Gamma_1 M_{21} - \Gamma_2 M_{12} }{ {\rm Tr}(\hat{\bf M}) } \right)^2 \right]  }
\end{equation}
is a $2\times 2$ symmetric matrix, and $\{M_{ij}\}$ represent the elements of the matrix $\hat{\bf M}$.  With a little algebra one can show that
\begin{equation}
  \hat{\bf X}\hat{\bf M} = \frac{ {\rm Tr}(\hat{\bf M}) }{B} ( A \hat{\bf \Gamma}^{-1} - \epsilon \hat{\bf Y} )\, ,
\label{eq:XM}
\end{equation}
where $A \equiv {\rm Tr}(\hat{\bf M})\det (\hat{\bf \Gamma})$, $\hat{\bf Y} \equiv \left( \begin{array}{cc} 0 & 1 \\ -1 & 0 \end{array} \right)$, $B \equiv \det (A \hat{\bf \Gamma}^{-1} - \epsilon \hat{\bf Y}) = A^2/\det(\hat{\bf \Gamma}) + \epsilon^2$, and $\epsilon \equiv \Gamma_1 M_{21} - \Gamma_2 M_{12} = 2 k_B R_1 R_2 C_c (T_2 - T_1)$.

Equation~\ref{eq:XM} can be rewritten as
\begin{equation}
  \hat{\bf X}\hat{\bf M} = \frac{ A' \hat{\bf \Gamma}^{-1} - \epsilon' \hat{\bf Y} }{A'(1+{\epsilon'}^2)} \, ,
\label{eq:XM2}
\end{equation}
where $A' \equiv \sqrt{\det \hat{\bf \Gamma}}$, and $\epsilon' \equiv \epsilon /[A' \cdot {\rm Tr}(\hat{\bf M})]$ gives a dimensionless measure for the deviation from thermal equilibrium.  Thus one has
\begin{equation}
  \hat{\bf{M}}^T \hat{\bf X} \hat{\bf{M}} = \left( \frac{\hat{\bf C}}{1+{\epsilon'}^2} \right) \left( \hat{\bf T}^{-1} - \frac{\epsilon' }{A'} \hat{\bf R}\hat{\bf Y}  \right) \, .
\end{equation}
Note that $\epsilon' = 0$ when thermal equilibrium between the two resistors is reached ($T_1=T_2$).  In this case, the matrix $\hat{\bf{M}}^T \hat{\bf X} \hat{\bf{M}} = \hat{\bf C} \hat{\bf T}^{-1}$, while the matrix $\hat{\bf T}$ reduces to a multiple of the identity matrix.  Therefore, $P_{\rm ss}$ exhibits a Boltzmann distribution and follows the shape of the equipotential contour.

The probability flux of the system is
\begin{equation}
  \vec{J} = - \hat{\bf M}^{-1} \vec{V} P - \frac12 \hat{\bf{M}}^{-1} \hat{\bf{\Gamma}} (\hat{\bf{M}}^{-1})^{T} \nabla P(\vec{V},t)\, .
\label{eq:j}
\end{equation}
in which the first term results from the restoring force towards the origin, while the second term can be attributed to the diffusive driving force.  At the steady state, one has
\begin{equation}
  \vec{J}_{\rm ss} = - \hat{\bf M}^{-1} [\vec{V} - \hat{\bf \Gamma} \hat{\bf X} \hat{\bf M}\vec{V} ]P_{\rm ss} \equiv \hat{\bf \Pi} \vec{V} P_{\rm ss} \,.
\label{eq:j_ss}
\end{equation}

In the case of thermal equilibrium, $T_1 = T_2$, Eq.~\ref{eq:XM2} reduces to $\hat{\bf X} \hat{\bf M} = \hat{\bf \Gamma}^{-1} $. 
As a result, $\vec{J}_{\rm ss} = \vec{0}$, which is a signature of the detailed balance.  On the other hand, in a NESS case, the two representative forces in Eq.~\ref{eq:j} do not cancel out, causing a persistent net flux in a NESS. The steady-state flux described by Eq.~\ref{eq:j_ss} is plotted as a vector field in Fig.~\ref{fig:figure2}(c).

\begin{figure*}
\includegraphics[width=160mm]{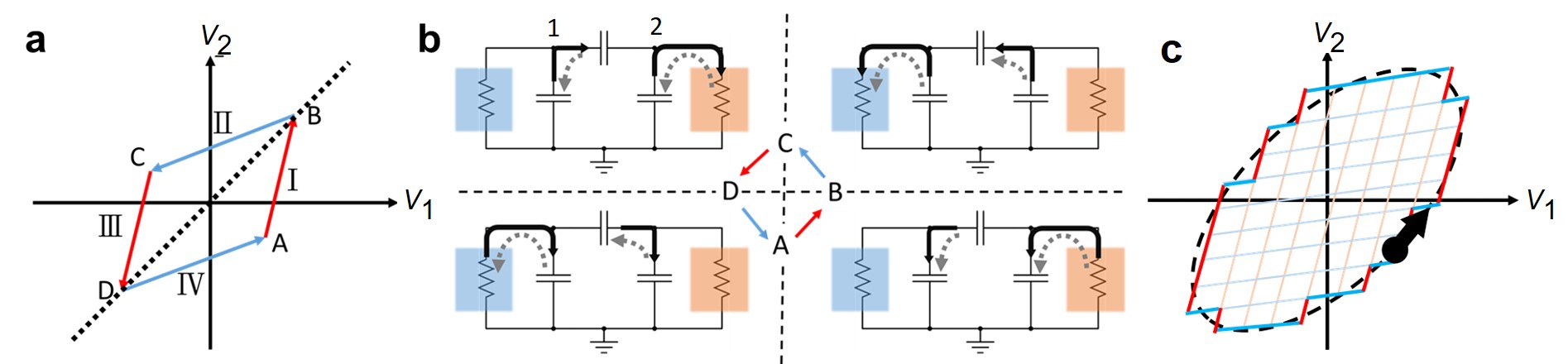}
\caption{Energy flow in a semi-adiabatic cycle.  (a) A closed cycle in the phase space formed by four chosen semi-adiabatic paths (I, II, III, IV) about the origin.  The red (blue) arrows indicate the processes adiabatic to $T_1$ ($T_2$), while the system is subject to energy exchanges with the $T_2$ ($T_1$) heat bath only.  (b) Major current flows (black solid arrows) and energy flows (grey dashed arrows) in the circuit for the four processes.  (c)An illustrated example of paving a closed cycle by semi-adiabatic processes.}
\label{fig:figure3}
\end{figure*}

The gyrating direction of the virtual particle can be identified by comparing the directions of $\vec{J}_{\rm ss}$ and $\nabla P_{\rm ss} \times \vec{z}$, where $\vec{z}$ is the unit vector point out of the $V_1 - V_2$ plane.  The virtual particle is gyrating counterclockwise if both vectors are parallel and clockwise if they are antiparallel.  Note that
\begin{align}
  &\vec{J}_{\rm ss} \cdot ( \nabla P_{\rm ss} \times \vec{z}) \!\! = - \nabla P_{\rm ss} \cdot \hat{\bf Y} \vec{J}_{\rm ss} \nonumber \\
  & \ \ = -\frac{2\epsilon}{{\rm Tr}(\hat{\bf M})} ( \hat{\bf X} \hat{\bf M} \vec{V})^{T} ( \hat{\bf{M}} \hat{\bf Y} \hat{\bf M}^{-1} \hat{\bf Y} ) (\hat{\bf X} \hat{\bf M} \vec{V}) P_{\rm ss}^2 \, .
\label{eq:JXgradP}
\end{align}
One can show that the matrix $\hat{\bf{M}} \hat{\bf Y} \hat{\bf M}^{-1} \hat{\bf Y}$ is symmetric, while ${\rm Tr}(\hat{\bf{M}} \hat{\bf Y} \hat{\bf M}^{-1} \hat{\bf Y})<0$ and $\det (\hat{\bf{M}} \hat{\bf Y} \hat{\bf M}^{-1} \hat{\bf Y})=1$.  As a result, $\hat{\bf{M}} \hat{\bf Y} \hat{\bf M}^{-1} \hat{\bf Y}$ is negative-definite, and on average, the virtual particle is gyrating counterclockwise if $\epsilon > 0$, i.e., $T_2 > T_1$, and vice versa.  As the direction of gyration coincides with that of heat transport from the hot towards the cold heat bath (see next section), we have therefore shown that for this coupled $RC$ circuit, on average, heat is transferred from the high- to low-temperature thermal baths.  Hence the second law of thermodynamics is recapitulated here through the discussion of gyrating dynamics.

\section{Energy flow in a closed trajectory}
\label{sec:cycle}
The above observation reveals that on average, the virtual particle rotates about the origin in the $V_1-V_2$ phase space, while its Brownian-motion signature is revealed within short-time intervals.  The average circulating behavior shows periodic oscillations in $V_1$ and $V_2$ with an identical frequency and a constant phase difference.  The system therefore acts like a mini electricity generator powered solely by the temperature difference, and thus can be compared to a Brownian ratchet or a mini heat engine.  The ac voltage could be conceivably used to power up devices or rectified to store electric energy.

How does energy transfer from the hot heat reservoir to the cold one in a directed cycle?  First we note that if the virtual particle circulates along some closed loop on the $V_1-V_2$ diagram, the amount of energy flowing into the circuit through resistor $R_i$ (the same as the amount of heat flowing out of the heat bath coupled to $R_i$) during one cycle is $Q_i = \oint V_i i_{R,i} dt = \oint V_i dq_i$, where $i_{R,i}$ is the current through $R_i$. One can find $Q_1 = -C_c \oint V_1 dV_2$ and $Q_2 = -C_c \oint V_2 dV_1 = C_c \oint V_1 dV_2 = -Q_1$, and their magnitude is proportional to the loop area $|\oint V_2 dV_1|$ on the $V_1-V_2$ diagram.  Thus over each counterclockwise cycle $Q_2 = -Q_1 > 0$, and a net energy is flowing from the hot reservoir towards the cold one.

The energy transfer can be better understood using Fig.~\ref{fig:figure3}(a) as a schematic example.  The cyclic diagram is constituted by four simple paths (I, II, III, IV), which form a parallelogram with endpoints A, B, C, and D.  The four paths are chosen such that $q_1$ stays fixed during I and III while $q_2$ stays fixed during II and IV.  As a consequence, for processes I and III, no currents are flowing through $R_1$, and the system can be considered adiabatic to the cold reservoir $T_1$, while it can exchange energy with the hot reservoir $T_2$.  Similarly, for processes II and IV, $i_{R,2}=0$, and the system can be considered adiabatic to the $T_2$ bath, while it can exchange energy with  the $T_1$ reservoir.

One can show that during processes I and III, the resistor $R_2$ is exerting positive work on capacitors $C_1$ and $C_2$, while the bridging capacitor $C_c$ is discharging and also releasing energy into the other capacitors.  And during processes II and IV, the capacitors $C_1$ and $C_2$ are discharging and releasing energy into $R_1$ and $C_c$.  The directions of net energy flows and electric currents are shown in Fig.~\ref{fig:figure3}(b).  Other than the reversed polarity in charges and currents, the processes III and IV simply repeat I and II, respectively.  After a full cycle, the system resumes its original state, and a net energy is transferred from the $T_2$ to the $T_1$ heat bath through the circuit elements.  The amount of transferred energy can be characterized via the enclosed area of the cycle, as larger cycles and faster gyrating rates signify higher heat conduction rates.

Note that for a parallelogram of the aforementioned semi-adiabatic processes without centering at the origin, the magnitude and even the sign of transported energy to and from the capacitors for each individual process may vary.  Yet the total amount of energy transported in the two processes I and III remains unchanged (and so on for II and IV).  As a result, the energy transfer can be characterized in terms of area on the $V_1-V_2$ plot, i.e., $|\oint V_2 dV_1|$.  Furthermore, one can dissect any closed cycle (e.g. the elliptical contour in our experimental observation in a NESS) into infinite pavements of parallelograms (see Fig.~\ref{fig:figure3}(c) for an illustrated example).  Thus any closed cycle can be treated as a composite of semi-adiabatic processes.

The linear coupled circuit described here does not convert any heat into work.  Therefore, although the gyrating behavior is observed in our system, currently it remains meaningless to discuss about its efficiency and output power.  Nonetheless, we can briefly remark on the possibility of extracting work from the system.  For all our discussed cases, the average entropy of this stochastic system (up to addition by some constant owing to $P_{\rm ss}$ is not a dimensionless quantity) is
$\langle S\rangle / k_B = - \displaystyle\int P_{\rm ss}(\vec{V})\ln P_{\rm ss}(\vec{V}) d\vec{V} =  \langle \vec{V}^T (\hat{\bf{M}}^T \hat{\bf X} \hat{\bf{M}}) \vec{V} \rangle + \frac12 \ln \frac{\pi^2}{\det (\hat{\bf{M}}^T \hat{\bf X} \hat{\bf{M}})} \nonumber  = 1+\ln (2 k_B \pi) - \frac{1}{2}\ln \det(\hat{\bf C}) + \frac{1}{2}\ln \left\{ T_1 T_2 + \frac{ C_c^2 R_1 R_2 (T_2-T_1)^2}{[\rm Tr (\hat{\bf R} \hat{\bf C})]^2} \right\}$.  Furthermore, the average internal energy is $\langle U\rangle = \frac {1}{2}k_B (T_1+T_2)$.  Therefore, for the NESS case we consider, where the circuit is thermalized by two different heat baths at $T_1$ and $T_2$, its average energy is identical with that in thermal equilibrium with the mean temperature $T= (T_1 + T_2)/2$.  On the other hand, one can easily show that the NESS average entropy is less than the equilibrium result at the average temperature, suggesting that the circuit in a NESS is more ordered.  Since in thermal equilibrium, entropy is a monotonic function of energy, thus in principle, the circuit in a NESS should be capable of providing work via some relaxation process towards the equilibrium where the system entropy is preserved.  Note that with proper external driving, this linear system can function as a heat engine or a refrigerator \cite{Park16}.

\section{Rotation speed of the gyrator}
\label{sec:rotation_speed}

\begin{figure}
\includegraphics[width=85mm]{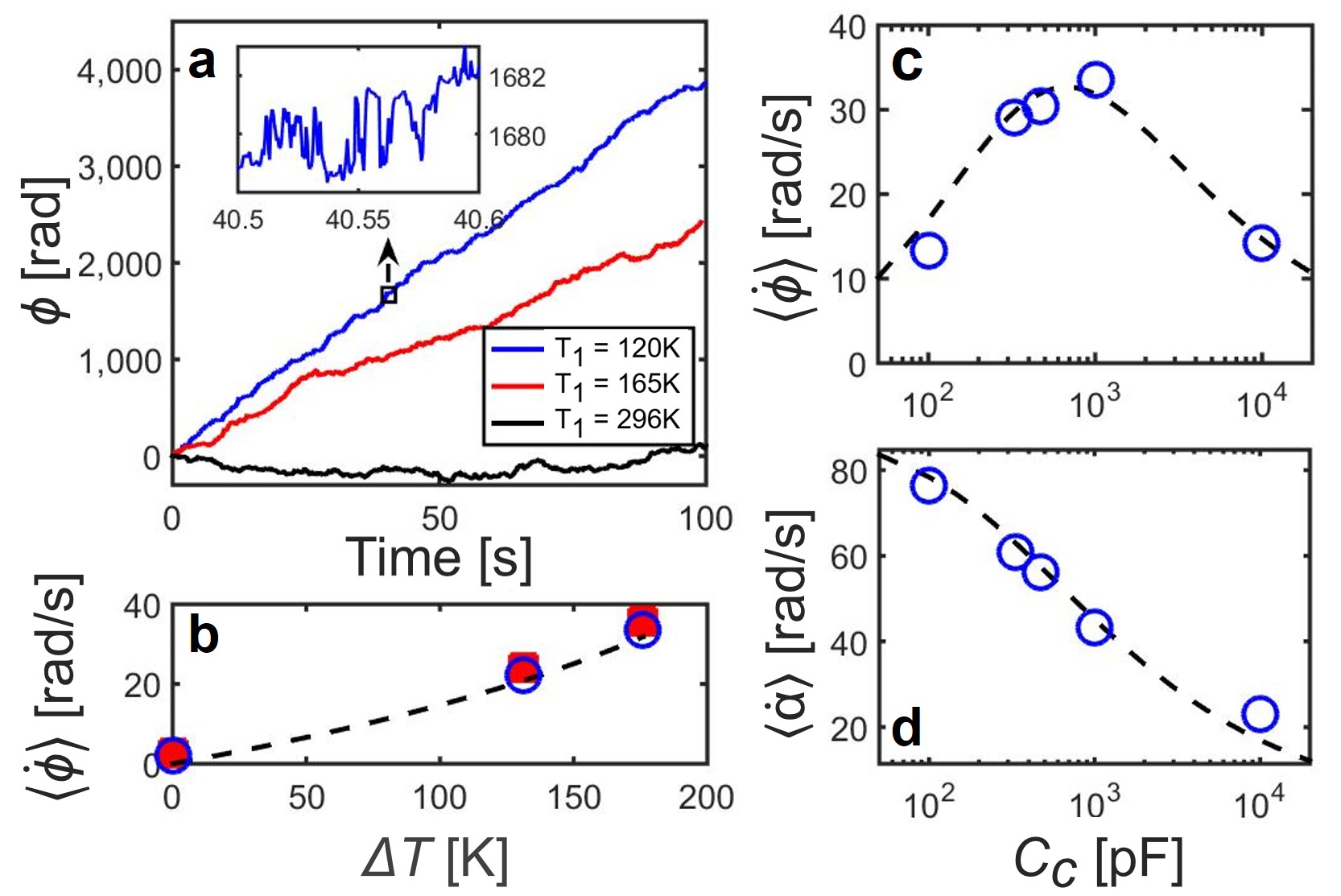}
\caption{Rotation speed of gyrator. (a) The measured time evolution in the angle of the virtual Brownian particle position $\phi(t)$ for various $T_1$ ($C_c$ = 1.0 nF being used).  The inset provides a zoom-in detail for fluctuating motions. (b) $\langle\dot\phi\rangle$ vs. $\Delta T$ evaluated from the average slope of $\phi(t)$ (solid square) and the average rotating speed derived from $\vec{v}_{\rm flow}$ (open circle) ($C_c$ = 1.0 nF being used).  Theoretical result is provided as the dashed curve. (c) $\langle\dot\phi\rangle$ vs. $C_c$ for $T_1=$ 120 K.  Symbols show the results from measurement while the dashed curve gives the theoretical prediction. (d) The averaged leading angle $\langle \alpha \rangle$, defined by the advanced phase of oscillation in $V_1$ over that in $V_2$, for a virtual particle circulating along the flux field ($T_1$ = 120 K; the same symbols as in (c)).}
\label{fig:figure4}
\end{figure}

The gyrating motion of the virtual Brownian particle can also be well visualized through the dynamics of $\displaystyle \phi(t) \equiv \tan^{-1}\left[\frac{V_2(t)}{V_1(t)}\right]$ \cite{defPhi}, the angle of the particle position vector $\vec{V}(t)$ relative to the $V_1$ axis in the configuration space. Fig.~\ref{fig:figure4}(a) presents the gross behavior of $\phi(t)$ for various $T_1$, while the stochastic behavior is magnified in the inset.  In the NESS cases ($T_1 < T_2$), the overall trend exhibits a linear growth in time, while such feature is absent in the thermal-equilibrium case ($T_1=T_2=$ 296 K).

The dependence of average gyrating rate, $\langle\dot\phi\rangle$, on the temperature difference, $\Delta T \equiv T_2-T_1$, is shown in Fig.~\ref{fig:figure4}(b).  Note that experimentally we obtain $\langle\dot\phi\rangle$ via two methods: The first method is finding the slopes of the fitted straight lines in Fig.~\ref{fig:figure4}(a) (solid square), while the second is evaluating the average rotating speed from the probability flux: $\langle\dot\phi\rangle = \displaystyle\int \! \frac{\vec{V} \times \vec{v}_{\rm flow}}{V^2} P_{\rm ss} d\vec{V}$ (open circle).  Both experimental evaluations agree well and indicate an approximately proportional relation between $\langle\dot\phi\rangle$ and $\Delta T$.

The temperature dependence on the gyrating flux is also studied analytically. By defining $\hat{\bf \Delta} \hat{\bf M} \equiv \hat{\bf \Gamma} \hat{\bf X} \hat{\bf M}  - \hat{\bf I}$, where $\hat{\bf I}$ is the $2\times 2$ identity matrix, one can show that
\begin{equation}
  \hat{\bf \Delta} \hat{\bf M} = \frac{-\epsilon^2 \hat{\bf I} - \epsilon {\rm Tr}(\hat{\bf M}) \hat{\bf \Gamma} \hat{\bf Y}}{B}
  = - \frac{\epsilon \hat{\bf Y} \hat{\bf X} \hat{\bf M} }{{\rm Tr}(\hat{\bf M})}\, .
\label{eq:DeltaM}
\end{equation}
Therefore, $\hat{\bf \Pi} = \hat{\bf M}^{-1} \hat{\bf \Delta} \hat{\bf M}$ (refer to Eq. (\ref{eq:j_ss})), and $\hat{\bf \Pi}$, $\vec{J}_{\rm ss}$ and thus $\langle \dot{\phi} \rangle$ are approximately proportional to $\epsilon$ and hence $\Delta T$ when the temperature difference is small.  The theoretical prediction of this linear behavior is shown by the dashed line in Fig.~\ref{fig:figure4}(b), as observed experimentally.

\par
We further study the dependence of rotating speed on the coupling, $C_c$, as is shown in Fig.~\ref{fig:figure4}(c).  Remarkably, $\langle\dot\phi\rangle$ does not increase monotonically with $C_c$.  Our theoretical result (dashed curve in Fig.~\ref{fig:figure4}(c)) predicts a broad peak near $C_c \approx$  700 pF, while the peak circulating speed is about 5 rev/s.  And the evaluation of $\langle\dot\phi\rangle$ from the experimental data (open circles) follow well with the theoretical curve, proving the existence of an optimal coupling for gyrating.

To investigate how the rotational speed of the virtual particle relates to the coupling, we first find from the average heat transfer rate \cite{Ciliberto13}
\begin{equation}
  \langle\dot {Q}\rangle = \frac{C_c^2 k_B(T_2-T_1)}{\det(\hat{\bf C})\cdot{\rm Tr}(\hat{\bf R}\hat{\bf C})} \, .
\label{eq:heatrate}
\end{equation} that $\langle \dot{Q} \rangle $ increases monotonically over $C_c$, as $\langle \dot{Q} \rangle \sim O(C_c^2)$ in the weak-coupling regime (when $C_c$ is small) and $\langle \dot{Q} \rangle \sim O(1)$ in the strong coupling regime (large $C_c$).  Furthermore, the average heat transfered from $T_2$ to $T_1$ reservoir during one gyrating cycle, $Q_{\rm cycle}$, is equal to the product of $C_c$ and the average area of gyration on the $V_1-V_2$ diagram.  Therefore, $Q_{\rm cycle}$ is proportional to $ C_c \pi/ \sqrt{\det (\hat{\bf M}^T \hat{\bf X} \hat{\bf M})}$.  One can show that $Q_{\rm cycle}$ is also increasing monotonically over $C_c$, as $Q_{\rm cycle} \sim O(\sqrt{C_c})$ for large $C_c$ and $Q_{\rm cycle} \sim O(C_c)$ for small $C_c$.  As a result, $\langle \dot{\phi} \rangle \approx 2\pi \cdot \langle \dot{Q} \rangle / Q_{\rm cycle}$ leads to $\langle \dot{\phi}\rangle \sim O(C_c)$ for small $C_c$ and $\langle \dot{\phi}\rangle \sim O(C_c^{-1/2})$ for large $C_c$.  Hence $\langle \dot{\phi}\rangle$ is not increasing monotonically; instead it reaches a peak, as evidenced in Fig.~\ref{fig:figure4}(c). The decreasing trend of $\langle\dot\phi\rangle$ can also be understood by recognizing that at large $C_c$ it takes a long time for the system to charge/discharge.

In Fig.~\ref{fig:figure4}(d) we present the average phase difference $\langle \alpha \rangle$ between $V_1$ and $V_2$ along the elliptical contours of constant $P_{\rm ss}(\vec{V})$ (positive if $V_1$ leads $V_2$). The leading angle $\langle \alpha \rangle$ is experimentally evaluated by the average of the instantaneous angle difference $\displaystyle \alpha=\tan^{-1}\left(\frac{-\dot{V_1}}{\omega V_1}\right) - \tan^{-1}\left(\frac{-\dot{V_2}}{\omega V_2}\right)$, where $\displaystyle \omega=\left|\frac{\vec{V}\times \dot{\vec{V}}}{V^2}\right|$ is the instantaneous angular velocity of the virtual particle in the $V_1-V_2$ phase space. For the special case that $C_c$ vanishes, the elliptical contours are nontilted, and thus $\langle \alpha \rangle$ is equal to 90 degrees.  As $C_c$ increases, the ellipses start to tilt due to the coupling between the signals $V_1$ and $V_2$, and as a result $\langle \alpha \rangle$ decreases.  Again the experimental results are well confirmed by theoretical analysis.

\section{Conclusion}
We demonstrate in this work that the linear, coupled {\it RC} circuit system, under the agitation of two different thermal baths near room temperature scale, can serve as a non-mechanical realization of the autonomous Brownian gyrator.  The incomplete cancelation between the diffusive drive and the potential-gradient dragging accounts for the net circulating flux in the steady state.  For such an arrangement, the system acts like a mini electricity generator, while the possibility for the usage of this generated power is still being explored.

The observation that heat is conducted from the hot to the cold reservoir is simply consistent with the second law of thermodynamics.  Yet the heat-transfer mechanism through the gyration in the configuration space is plausible, noting that the conducting element possesses two thermal degrees of freedom only.  The direction of heat flow, the gyrating dynamics, and the total entropy production, are all representations of the second law, which speaks of time irreversibility in a nonequilibrium steady state.  Our study helps re-affirm the general principle and the possible realization of a Brownian ratchet under real thermal baths.

%

\begin{acknowledgments}

The authors wish to thank Yonggun Jun, Jae Dong Noh, and Rafael Sanchez for helpful comments.  This work has been supported by Ministry of Science and Technology in Taiwan under grant MOST 104-2112-M-008-003-MY3, 105-2112-M-008-019-MY2, and 105-2112-M-008-023.  C.-L.L. acknowledges the support from NCTS thematic group Complex Systems.

\end{acknowledgments}


\begin{thebibliography}{0}

 \bibitem{Feynman63}
R. P. Feynman, R. B. Leighton, and M. Sands, \textit{The Feynman Lectures on Physics, Vol. 1} (Addison-Wesley, Boston, 1963).

 \bibitem{Martinez17}
I. A. Martinez \textit{et al.}, Soft Matter \textbf{13}, 22 (2017).

 \bibitem{Reimann02}
P. Reimann, Phys. Rep. \textbf{361}, 57 (2002).

 \bibitem{Hanggi09}
P. H{\"a}nggi and F. Marchesoni, Rev. Mod. Phys. \textbf{81}, 387 (2009).

 \bibitem{VanDenBroeck04}
C. Van den Broeck, R. Kawai, and P. Meurs, Phys. Rev. Lett. \textbf{93}, 090601 (2004).

 \bibitem{Steeneken11}
P. G. Steeneken \textit{et al.}, Nat. Phys. \textbf{7}, 354 (2011)

 \bibitem{Blickle12}
V. Blickle and C. Bechinger, Nat. Phys. \textbf{8}, 143 (2012).

 \bibitem{Martines16}
I. A. Martinez \textit{et al.}, Nat. Phys. \textbf{12}, 67 (2016).

 \bibitem{Koski16}
J. V. Koski, A. Kutvonen, I. M. Khaymovich, T. Ala-Nissila and J.P. Pekola,  Phys. Rev. Lett. \textbf{115}, 260602 (2015).

\bibitem{Davis01}
B. R. Davis, D. Abbott, and J. M. R. Parrondo, Chaos \textbf{11}, 747 (2001).

 \bibitem{Filliger07}
R. Filliger and P. Reimann, Phys. Rev. Lett. \textbf{99}, 230602 (2007).

 \bibitem{Kwon05}
C. Kwon, P. Ao, and D. J. Thouless, Proc. Natl. Acad. Sci. \textbf{102}, 13029 (2005).

\bibitem{Ghanta17}
A. Ghanta, J. C. Neu and S. Teitsworth, Phys. Rev. E  \textbf{95}, 032128 (2017).

 \bibitem{Seifert12}
U. Seifert, Rep. Prog. Phys. \textbf{75}, 126001 (2012).

 \bibitem{Eshuis10}
P. Eshuis, K. van der Weele, D. Lohse and D. van der Meer, Phys. Rev. Lett. \textbf{104}, 248001 (2010).

 \bibitem{Peng16}
Z. Peng and K. To, Phys. Rev. E \textbf{94}, 022902 (2016).

 \bibitem{Serra-Garcia16}
M. Serra-Garcia \textit{et al.}, Phys. Rev. Lett. \textbf{117}, 010602 (2016).

\bibitem{Hartmann15}
F. Hartmann, P. Pfeffer, S. H{\"o}fling, M. Kamp and L. Worschech, Phys. Rev. Lett. \textbf{114}, 146805 (2015).

 \bibitem{Sanchez11}
R. S{\'a}nchez and M. B{\"u}ttiker, Phys. Rev. B \textbf{83}, 085428 (2011).

 \bibitem{Sothmann12}
B. Sothmann, R. S{\'a}nchez, A.N. Jordan and M. B{\"u}ttiker, Phys. Rev. B \textbf{85}, 205301 (2012).

 \bibitem{Whitney16}
R. S. Whitney \textit{et al.}, Physica E \textbf{75}, 257 (2016).

 \bibitem{Roche15}
B. Roche \textit{et al.}, Nat. Commun. \textbf{6}, 6738 (2015).

 \bibitem{Thierschmann15}
H. Thierschmann \textit{et al.}, Nat. Nanotechnol. \textbf{10}, 854 (2015).

 \bibitem{Ciliberto13}
S. Ciliberto, A. Imparato, A. Naert and M. Tanase, Phys. Rev. Lett. \textbf{110}, 180601 (2013).

 \bibitem{Chiang17}
K.-H. Chiang,  C.-L. Lee, P.-Y. Lai and Y.-F. Chen, Phys. Rev. E \textbf{95}, 012158 (2017).


\bibitem{JpGP}
As an alternative proof, first note that $\nabla P_{\rm ss} = -2 \hat{\bf{M}}^T \hat{\bf X} \hat{\bf{M}} \vec{V} P_{\rm ss}$.  According to Eqs.~(\ref{eq:j_ss}) and (\ref{eq:DeltaM}),
$ \displaystyle
\nabla P_{\rm ss} \cdot \vec{J}_{\rm ss}  = -2 \vec{V}^{T} (\hat{\bf{M}}^T \hat{\bf X} \hat{\bf{M}}) (\hat{\bf M}^{-1} \hat{\bf \Delta} \hat{\bf M}) \vec{V} P_{\rm ss}^2
 = \frac{2\epsilon}{{\rm Tr}(\hat{\bf M})} \vec{V}^{T} (\hat{\bf X}\hat{\bf{M}})^T \hat{\bf Y} (\hat{\bf X}\hat{\bf M}) \vec{V} P_{\rm ss}^2 =0
$
for all $\vec{V}$, since the matrix $(\hat{\bf X}\hat{\bf{M}})^T \hat{\bf Y} (\hat{\bf X}\hat{\bf M})$ is antisymmetric.

 \bibitem{Park16}
J.-M. Park, H.-M. Chun, and J. D. Noh, Phys. Rev. E \textbf{94}, 012127 (2016).

\bibitem{defPhi}
The time evolution of $\phi(t)$ is evaluated assuming minor change of $\vec{V}(t)$ in the phase space, i.e, $-\pi < \phi(t+\Delta t)- \phi(t) \leq \pi$. The argument is based on that the sampling interval $\Delta t \sim$ 0.5 ms is faster than the autocorrelation time ( few milliseconds ) of $\vec{V}(t)$.

\end{thebibliography}
\end{document}